\documentclass[prl,aps,twocolumn,showpacs,amsmath,amssymb,floatfix]{revtex4}

\usepackage[dvips]{graphicx}
\usepackage{bm}
\newcommand{\rot}{\bm\nabla\wedge}

\begin{document}

\title{Fermi Gases in Slowly Rotating Traps:
Superfluid vs Collisional Hydrodynamics}
\author{Marco Cozzini}
\author{Sandro Stringari}
\affiliation{Dipartimento di Fisica, Universit\`a di Trento and BEC-INFM, I-38050 Povo, Italy}
  
\date{\today}

\begin{abstract}
The dynamic behavior of a Fermi gas confined in a deformed trap rotating at
low angular velocity is investigated in the framework of hydrodynamic theory.
The differences exhibited by a normal gas in the collisional regime and a
superfluid are discussed. Special emphasis is given to the collective
oscillations excited when the deformation of the rotating trap is suddenly
removed or when the rotation is suddenly stopped. The presence of vorticity in
the normal phase is shown to give rise to precession and beating phenomena
which are absent in the superfluid phase.
\end{abstract}

\pacs{03.75.Ss, 67.40.Db, 03.75.Kk}

\maketitle

Recent experiments on ultracold atomic Fermi gases close to a Feshbach
resonance \cite{thomas,jin,salomon} have explicitly revealed the emergence of a
hydrodynamic regime which shows up in the anisotropic shape of the expanding
gas. 
This behavior is the consequence of the anisotropy of the pressure gradients
which are stronger in the tighter directions of the trap. It should be
contrasted with the isotropic nature of the expansion of the non interacting
Fermi gas, which instead reflects the initial isotropy of the momentum
distribution. 
In Ref.~\cite{menotti} the anisotropy of the expansion of a Fermi gas has been
suggested as a possible signature of superfluidity.
The analysis of Ref.~\cite{menotti} concerns, however, dilute gases far from
Feshbach resonances where the system, in the normal phase, is collisionless at
low temperature due to Pauli blocking \cite{vichi1} and consequently expands
differently from a superfluid.
If one instead works close to a Feshbach resonance the huge increase of the
collisional rate will favor the achievement of the hydrodynamic regime in the
normal phase even at low temperature \cite{thomas}.
At the same time the resonance effect is expected to enhance significantly the
value of the critical temperature for superfluidity \cite{holland}, providing
promising perspectives for its experimental realization. 
Since the anisotropy of the expansion is compatible with either the collisional
and superfluid hydrodynamic pictures, its experimental observation cannot be
used as a test of superfluidity without further considerations. As a
consequence, it is important to identify alternative effects which permit to
distinguish between the two regimes.

The purpose of the present letter is to show that the study of the collective
oscillations of a trapped Fermi gas rotating at low angular velocity can
provide a useful identification of superfluidity.
It is in fact well known that a superfluid cannot support vorticity, unless
quantized vortices are created. In the following we will consider situations
where vortices are absent. This can be ensured by rotating the confining trap
at sufficiently low angular velocities. Under these conditions the dynamic
behavior of a superfluid is described by the equations of irrotational 
hydrodynamics:
\begin{equation} \label{eq:cont}
\frac{\partial{n}}{\partial t}+\bm\nabla\cdot({n}\bm{v}) = 0 \ ,
\end{equation}
\begin{equation} \label{eq:euler irrHD}
\frac{\partial\bm{v}}{\partial t}\!=\!
-\!\bm\nabla\!\left(\!\frac{v^2}{2}\!+\!\frac{V_{\text{ext}}}{M}\!+
\!\frac{\mu}{M}\!\right)\!=\!
-\!\bm\nabla\!\left(\!\frac{v^2}{2}\!+\!\frac{V_{\text{ext}}}{M}\!\right)\!
-\!\frac{\bm\nabla P}{M{n}} ,
\end{equation}
where $V_{\text{ext}}$ is the external potential generated by the confining
trap, while $P$ and $\mu$ are, respectively, the pressure and the chemical
potential of a uniform gas evaluated at the corresponding density. In the last
identity we have used the $T=0$ thermodynamic relationship
$\bm\nabla{P}={n}\bm\nabla\mu$. The above equations apply to dynamic phenomena
of macroscopic type where the local density approximation to the equation of
state is justified. They hold for both Bose and Fermi superfluids at zero
temperature. At finite temperature, below $T_c$, they should be generalized to
the equations of two fluid hydrodynamics (see, for example, \cite{landau}).
Equations (\ref{eq:cont}) and (\ref{eq:euler irrHD}) have been systematically
used in the last years to test the effects of superfluidity on the dynamic
behavior of Bose-Einstein condensed gases \cite{PS}. 
  
Differently from a superfluid, a normal gas can support vorticity and in the
collisional regime the equation for the velocity field takes the classical
Euler form
\begin{equation} \label{eq:euler rotHD}
\frac{\partial\bm{v}}{\partial t} =
-\bm\nabla\left(\frac{v^2}{2}+\frac{V_{\text{ext}}}{M}\right)-
\frac{\bm\nabla P}{M{n}}+\bm{v}\wedge(\rot\bm{v}) \ ,
\end{equation}
where the time dependence of the pressure should be calculated taking into
account the conditions of adiabaticity, following from the conservation of
entropy.
Equation (\ref{eq:euler rotHD}) differs from equation (\ref{eq:euler irrHD})
for the superfluid velocity because of the last term, proportional to the
vorticity $\rot\bm{v}$.
The hydrodynamic description (\ref{eq:euler rotHD}) holds provided the
collisional relaxation time $\tau$  is much smaller than the inverse of the
typical frequencies $\omega$ characterizing the dynamic phenomena under
investigation, fixed by the oscillator trapping frequencies: $\omega\tau\ll1$.

Let us first suppose that the initial state of the system does not contain any
velocity flow (gas at rest in a static trap).
In this case the equations for the expansion as well as for the linearized
collective oscillations take the same irrotational form both in the superfluid
and classical cases.
In the following we will assume a trapping potential of harmonic shape:
$V_{\text{ext}}=M\,(\omega_x^2\,x^2+\omega_y^2\,y^2+\omega_z^2\,z^2)/2$.
If the potential is axially symmetric ($\omega_x=\omega_y=\omega_\perp$), one
can classify the solutions for the linearized oscillations in terms of the
third component $\hbar m$ of angular momentum.
We will consider solutions where the velocity field is linear in the spatial
coordinates. The solutions with $m=\pm2$ and $m=\pm1$ are surface excitations
of the form $\bm{v}\propto\bm\nabla(x\pm iy)^2$ and
$\bm{v}\propto\bm\nabla[(x\pm iy)z]$ respectively, with frequencies given by
\begin{eqnarray}
\label{eq:m+-2 irr}\omega(m=\pm2) & = & \sqrt2\,\omega_\perp \ , \\
\label{eq:m+-1 irr}\omega(m=\pm1) & = & \sqrt{\omega_\perp^2+\omega_z^2} \ ,
\end{eqnarray}
independent of the equation of state.
Assuming a power law dependence for the isoentropic equation of state
($P\propto{n}^{\gamma+1}$), one easily finds analytic solutions also for the
$m=0$ modes which are characterized by a velocity field of the form
$\bm{v}\propto\bm\nabla[a(x^2+y^2)+bz^2]$.
The corresponding collective frequencies are given by
\begin{equation} \label{eq:m=0 axisymm gamma}
\begin{array}{l}
\displaystyle\omega^2(m=0) \,\ = \,\ \frac{1}{2}\,
\bigg[2\,(\gamma+1)\,\omega_\perp^2+(\gamma+2)\,\omega_z^2 \,\ \pm \\
\sqrt{4(\gamma+1)^2\omega_\perp^4+(\gamma+2)^2\omega_z^4
+4(\gamma^2-3\gamma-2)\omega_\perp^2\omega_z^2}\,\bigg] \ .
\end{array}
\end{equation}
Equation (\ref{eq:m=0 axisymm gamma}) reduces to the one derived in
Ref.~\cite{sandro} in the interacting Bose case ($\gamma=1$), while in the case
of the ideal gas ($\gamma=2/3$) it reduces to the predictions of
Refs.~\cite{griffin,minguzzi2,kagan} and, for spherical trapping, to the ones
of Refs.~\cite{baranov}.
For elongated traps ($\omega_z\ll\omega_\perp$) one finds
$\omega=\sqrt{2(\gamma+1)}\,\omega_\perp$ and
$\omega=\sqrt{(3\,\gamma+2)/(\gamma+1)}\,\omega_z$.

The analysis of the collective frequencies is easily generalized to tri-axial
anisotropy ($\omega_x\ne\omega_y\ne\omega_z$), where one finds 3 solutions of
the form $\bm{v}\propto\bm\nabla(xy)$, $\bm{v}\propto\bm\nabla(xz)$,
$\bm{v}\propto\bm\nabla(yz)$.
These are the so called scissors modes \cite{david scissors} relative to the
three pairs of axes, with frequency $\sqrt{\omega_x^2+\omega_y^2}$,
$\sqrt{\omega_x^2+\omega_z^2}$ and $\sqrt{\omega_y^2+\omega_z^2}$, respectively.
The other three solutions have the form
$\bm{v}\propto\bm\nabla(ax^2+by^2+cz^2)$ and their frequencies obey the
equation
\begin{equation} \label{eq:omega6 gamma}
\begin{array}{c}
\omega^6-(2+\gamma)(\omega_x^2+\omega_y^2+\omega_z^2)\,\omega^4+ \\
\rule[-0.3cm]{0cm}{0.8cm}+4\,(\gamma+1)
(\omega_x^2\,\omega_y^2+\omega_x^2\,\omega_z^2+
\omega_y^2\,\omega_z^2)\,\omega^2+ \\
-4\,(2+3\,\gamma)\,\omega_x^2\,\omega_y^2\,\omega_z^2 \,\ = \,\ 0 \ .
\end{array}
\end{equation}

Let us stress again that the above results hold both in the superfluid and
normal hydrodynamic phases.
In particular, if we excite the scissors mode by suddenly rotating the
confining trap starting from the ground state configuration, like in the
experiment of Ref.~\cite{foot}, the gas will oscillate with the same frequency
in both regimes.
The situation would be different in a dilute Fermi gas far from Feshbach
resonances where the dynamic response, in the normal phase, is collisionless at
low temperature and the behavior of the scissors oscillation exhibits
different features with respect to the superfluid case
\cite{david scissors,minguzzi}.
Although in a Fermi gas close to a Feshbach resonance the measurement of the
collective frequencies (\ref{eq:m+-2 irr}$-$\ref{eq:omega6 gamma}) around the
ground state does not provide a direct indication of superfluidity, their
experimental determination would be nevertheless very useful, providing an
accurate and quantitative proof of the achievement of the hydrodynamic regime.
This is best tested looking at the surface excitations, whose frequencies are
insensitive to the equation of state.

We are now ready to explore the rotational properties of the system. The most
natural procedure is to let a deformed trap rotate at angular velocity $\Omega$
in the $x$-$y$ plane.
The angular velocity should be turned on adiabatically in order to ensure the
conditions of stationarity.
Furthermore the final angular velocity $\Omega$ should be small enough to avoid
the formation of quantized vortices in the superfluid phase. This condition is
not very restrictive since the angular velocity needed to nucleate vortices by
adiabatic increase of the rotation rate is very high \cite{castin}, as
confirmed experimentally in the case of Bose-Einstein condensates
\cite{chevy nucleation}.

If the system is superfluid the stationary velocity field has the irrotational
form $\bm{v}=\alpha\bm\nabla(xy)$. The dependence of the coefficient $\alpha$
on the angular velocity $\Omega$ has been discussed in Ref.~\cite{recati} and
for small angular velocities reduces to $\alpha=-\epsilon\Omega$, where
$\epsilon=(\omega_x^2-\omega_y^2)/(\omega_x^2+\omega_y^2)$ is the deformation
of the trap in the $x$-$y$ plane.
Actually, in the limit of an axisymmetric trap ($\epsilon=0$), the velocity
field exactly vanishes revealing that the superfluid is not capable to rotate.
If instead the system is normal, the stationary velocity field takes the rigid
form $\bm{v}=\bm\Omega\wedge\bm{r}$, corresponding to
$\rot\bm{v}=2\,\bm\Omega$.
Notice that, while the rigid rotation is correctly described by the
hydrodynamic equations (\ref{eq:cont}) and (\ref{eq:euler rotHD}), the
thermalization process which permits its achievement starting from a gas
initially at rest, is not accounted for by Eq.~(\ref{eq:euler rotHD}), because
of the absence of viscosity terms.
The time needed to achieve the rigid rotation of the gas (spin-up time) was
calculated in Ref.~\cite{david} and in the hydrodynamic regime
$\omega_\perp\,\tau\ll1$ is of the order of
$(\epsilon^2\omega_\perp^2\tau)^{-1}$, where $\tau$ is the relaxation time
fixed by collisions and $\omega_\perp^2=(\omega_x^2+\omega_y^2)/2$. This
suggests that, in order to reach steady rigid rotation in reasonable times, the
system should not be too deeply in the hydrodynamic regime and at the same time
the deformation of the rotating trap should not be too small
\footnote{The presence of a residual, static anisotropy $\epsilon_{\text{st}}$,
that in general acts to despin the cloud even if
$\epsilon_{\text{st}}\ll\epsilon$, is not expected to play an important role in
the hydrodynamic regime, where the ratio between the spin-up and spin-down
times is given by $(\epsilon_{\text{st}}/\epsilon)^2$ \cite{david}.}.

The fact that the velocity field is so different in the two regimes, gives rise
to different predictions for the frequencies of the collective oscillations.
For an axisymmetric trap the differences become particularly clear. In fact,
while an axisymmetric superfluid cannot rotate and the frequencies of the
$m=\pm2$ quadrupole oscillations are given by Eq.~(\ref{eq:m+-2 irr}), in the
collisional hydrodynamic case the degeneracy of these modes is lifted by the
presence of the rigid rotation according to the law \cite{cozzini}:
\begin{equation} \label{eq:m+-2}
\omega(m=\pm2) =
\sqrt{2\,\omega_\perp^2-\Omega^2}\pm\Omega \ .
\end{equation}
Equation (\ref{eq:m+-2}) is consistent with the sum rule approach developed in
Ref.~\cite{zambelli}, yielding the result
\begin{equation} \label{eq:zambelli}
\Delta\omega = \omega(m=+2)-\omega(m=-2) = \frac{2}{M}\,
\frac{\langle l_z \rangle}{\langle x^2+y^2 \rangle}
\end{equation}
for the splitting of the $m=\pm2$ quadrupole frequencies, where
$\langle{l_z}\rangle$ is the angular momentum per particle.
In the case of a superfluid no angular momentum is carried by the system 
because of the irrotationality constraint, while in the collisional
hydrodynamic regime the angular momentum is given by the rigid value
$\langle{l_z}\rangle=M\,\Omega\langle{x^2+y^2}\rangle$ and hence
$\Delta\omega=2\,\Omega$.
Measuring the splitting $\Delta\omega$ then provides an efficient way to detect
the vorticity of the gas. 
The splitting can be measured by suddenly switching off the deformation of the
trap.
Immediately after, the system will feel the axisymmetric trap
$V_{\text{ext}}=M\,[\omega_\perp^2(x^2+y^2)+\omega_z^2\,z^2]/2$, but will no
longer be in equlibrium. Actually, in linear approximation the state of the
system can be written in the form
$|0\rangle+a_+|m=+2\rangle+a_-|m=-2\rangle$,
where $|0\rangle$ is the new equilibrium state, while $|m=\pm2\rangle$
are the $m=\pm2$ quadrupole states with excitation energies
$\hbar\omega_{\pm}\equiv\omega(m=\pm2)$.
The coefficients $a_{\pm}$ are fixed by the initial conditions, including the
quadrupole deformation and the angular velocity $\Omega$. One finds
$a_++a_-=\langle{x^2-y^2}\rangle/\sqrt\sigma$ and
$a_+\omega_+-a_-\omega_-=2\,\Omega\langle{x^2-y^2}\rangle/\sqrt\sigma$, where
$\sigma$ is the quadrupole strength \cite{zambelli}.
The time evolution of the states $|m=\pm2\rangle$ is fixed by the
frequencies $\omega_{\pm}$ and a simple calculation yields the result
\begin{equation} \label{eq:angle}
\tan(2\,\theta) =
\frac{b_+\tan(\Delta\omega\,t/2)+b_-\tan(\omega_Qt)}
{b_+-b_-\tan(\omega_Qt)\tan(\Delta\omega\,t/2)} \ ,
\end{equation}
where $\theta$ is the angle of the principal axis of the gas in the $x$-$y$
plane, $\omega_Q=(\omega_++\omega_-)/2$ and $b_\pm=a_+\pm a_-$.
In the superfluid case $\Delta\omega=0$ and one finds the result
$\tan(2\,\theta)=(2\,\Omega/\omega_Q)\tan(\omega_Qt)$. In the collisional
hydrodynamic case, the angle $\theta$ exhibits an additional slow precession.
This is best seen by taking stroboscopic images at times $t=2\pi n/\omega_Q$,
with $n$ integer, at which the deformation of the atomic cloud is maximum. For
such times $\tan(\omega_Qt)=0$ and Eq.~(\ref{eq:angle}) yields the precession
law $\theta=\Delta\omega\,t/4$.
This precession is caused by the splitting $\Delta\omega=2\,\Omega$ and is
absent in the superfluid case. 
The numerical solution of the hydrodynamic equations confirms (see Fig.1) the
accuracy of the prediction (\ref{eq:angle}), based on linear approximation.
In the numerical calculation we have chosen the value $\gamma=2/3$
characterizing the equation of state of an ideal gas, including the most
relevant case of a degenerate Fermi gas. The results are not however sensitive
to the value of $\gamma$, consistently with Eq.~(\ref{eq:angle}), unless large
values of $\epsilon$ are considered.
The proposed experiment is similar to the one used in \cite{chevy} to measure
the angular momentum of quantized vortices. 
In that experiment the deformation of the gas was produced by suddenly
switching on a laser field in an almost axy-symmetric Bose-Einstein
condensates. This corresponds to setting  $a_+=a_-$ ($b_-=0$) and
Eq.~(\ref{eq:angle}) reduces to $\theta=\Delta\omega\,t/4$.

\begin{figure}

\includegraphics[width=8.5cm]{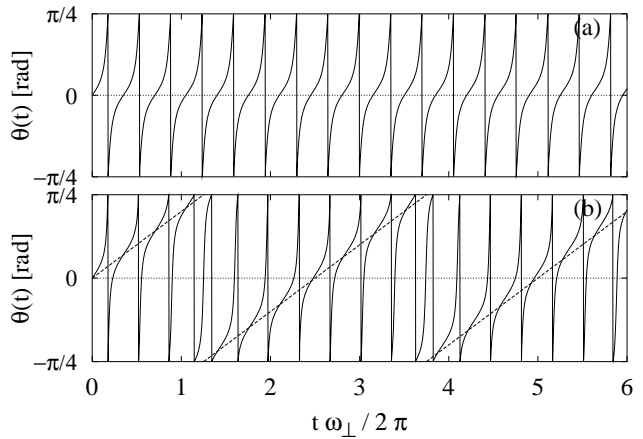}

\caption{\label{fig:angle eps=0}Time evolution of the angle $\theta$ (measured
in radians and plotted modulus $\pi/2$) after the sudden switch off of the trap
deformation.
(a) Superfluid hydrodynamics; (b) collisional hydrodynamics. The trap
parameters are $\epsilon=0.2$, $\Omega=0.2\,\omega_\perp$,
$\omega_z/\omega_\perp=0.1$, while $\gamma=2/3$. The dashed line in (b) is the
curve $\theta=\Omega\,t/2$ (see text).}

\end{figure}
 
\begin{figure}

\includegraphics[width=8.5cm]{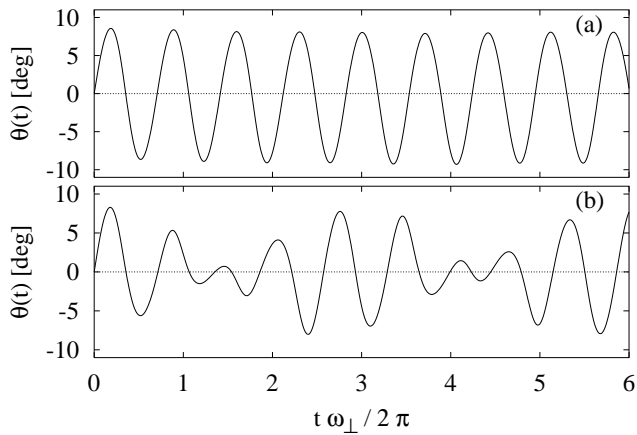}

\caption{\label{fig:comparison}Time evolution of the angle $\theta$ (measured
in degrees) after the sudden stop of the rotation of the trap.
(a) Superfluid hydrodynamics; (b) collisional hydrodynamics. Parameters as in
Fig.~\ref{fig:angle eps=0}.}

\end{figure}

If instead of switching off the deformation we simply stop the rotation of the
deformed trap, an other interesting phenomenon takes place that is worth
discussing.
In fact the gas, due to its inertia, will first continue rotating, but will
soon feel the restoring force produced by the deformed confining potential,
generating an oscillation around the new equilibrium configuration. In the
superfluid this procedure will excite the usual scissors mode with frequency
$\sqrt2\,\omega_\perp$.
In the case of classical hydrodynamics, the gas will instead oscillate
differently.
Actually a remarkable property exhibited by equations (\ref{eq:cont}) and
(\ref{eq:euler rotHD}) of classical hydrodynamics is that they admit stationary
solutions with non vanishing velocity flow also when the trap is at rest in the
laboratory frame.
These solutions have the form \cite{cozzini}
$\bm{v}=\bm\Omega\wedge\bm{r}+\alpha\bm\nabla(xy)$ and for small $\Omega$ one
finds $\alpha=\epsilon\Omega$.
Under the condition $\Omega\gg\epsilon^2\omega_\perp$, the resulting
oscillation can still be described as a linear combination of the two modes
(\ref{eq:m+-2}) and will consist of the characteristic beating
\footnote{In the opposite limit $\Omega\ll\epsilon^2\omega_\perp$, the beating
disappears and the gas oscillates with the frequency $\sqrt2\,\omega_\perp$.}
\begin{equation} \label{eq:theta}
\theta(t)=(\Omega/\omega_Q)\,\sin(\omega_Q\,t)\,\cos(\Delta\omega\,t/2) \ .
\end{equation}
Under the same conditions, also the intrinsic deformation $\delta$ of the cloud
exhibits an oscillatory behavior described by the law
$\delta(t)-\delta_0=-(2\,\Omega\,\epsilon/\omega_Q)\,
\sin(\omega_Q\,t)\,\sin(\Delta\omega\,t/2)$, where
$\delta_0=\langle{y^2-x^2}\rangle/\langle{x^2+y^2}\rangle=\epsilon$ is
the deformation of the stationary configuration.
In Fig.~\ref{fig:comparison} we compare the behavior of the angle $\theta$ in
the superfluid and normal cases. The numerical results have been obtained by
solving the hydrodynamic equations and in the collisional case exhibit the
typical beating predicted by Eq.~(\ref{eq:theta}).
	
Let us finally recall that the differences between the superfluid and
collisional regimes discussed in this letter concern genuine macroscopic
phenomena. The theory of superfluidity predicts also the occurrence of more
microscopic quantum phenomena, associated with the quantization of circulation
and the appearance of quantized vortices.
The description of quantized vortices requires theoretical schemes beyond the
hydrodynamic picture and has been the object of recent work in Fermi
superfluids in various regimes, including the BCS \cite{viverit} and the
unitarity limit \cite{bulgac}.
Their observation, similarly to the case of Bose-Einstein condensates
\cite{chevy}, could be again revealed by the splitting (\ref{eq:zambelli}) of
the quadrupole frequencies. In a Fermi superfluid, one predicts
$\langle{l_z}\rangle=\hbar/2$ for a single vortex line aligned along the
symmetry axis
\footnote{It is useful to compare the angular momentum of a single vortex line
with the rigid value predicted in the normal phase. For an ideal degenerate
Fermi gas in a harmonic trap one finds $\langle{l_z}\rangle=
(\hbar/2)(6\,N\omega_z/\omega_\perp)^{1/3}(\Omega/\omega_\perp)
[1-(\Omega/\omega_\perp)^2]^{-2/3}$, where $N$ is the number of atoms of each
spin component.}.
The realization of quantized vortices however requires different procedures
with respect to the ones discussed in the present work.
In particular one should likely work at higher angular velocity, close to
$\omega_\perp/\sqrt2$ where the superfluid becomes unstable against the
formation of quadrupole deformations \cite{recati,dalfovo}. Furthermore, one
should switch on the rotation of the trap in a non adiabatic way in order to
favor their nucleation.
This procedure has already proven to be successful in the experimental
realization of quantized vortices in Bose-Einstein condensates \cite{ENS}.

\begin{acknowledgments}
Stimulating discussions with L.~P.~Pitaevskii are acknowledged.
\end{acknowledgments}


\begin{thebibliography}{99}

\bibitem{thomas}
K.~M.~O'Hara \textit{et al.}, Science \textbf{298}, 2179 (2002).  
  
\bibitem{jin}
C.~A.~Regal and D.~S.~Jin, cond-mat/0302246.

\bibitem{salomon}
T.~Bourdel \textit{et al.}, cond-mat/0303079.

\bibitem{menotti}
C.~Menotti, P.~Pedri, and S.~Stringari, Phys. Rev. Lett. \textbf{89}, 250402
(2002).

\bibitem{vichi1}
L.~Vichi, J. Low. Temp. Phys. \textbf{121}, 177 (2000).

\bibitem{holland}
M.~Holland \textit{et al.}, Phys. Rev. Lett. \textbf{87}, 120406 (2001); 
E.~Timmermans \textit{et al.}, Phys. Lett. A \textbf{285}, 228 (2001);
Y.~Ohashi and A.~Griffin, Phys. Rev. Lett. \textbf{89}, 130402 (2002).

\bibitem{landau} L.~D.~Landau and E.~M.~Lifshitz, \textit{Fluid Mechanics} (2nd
edn), Pergamon Press, Oxford (1987).
 
\bibitem{PS}
L.~P.~Pitaevskii and S.~Stringari, \textit{Bose-Einstein Condensation},
Oxford University Press (2003);
F.~Dalfovo \textit{et al.}, Rev. Mod. Phys. \textbf{71}, 463 (1999).
	
\bibitem{chevy next} F.~Chevy and S.~Stringari, to be published.

\bibitem{sandro}
S.~Stringari, Phys. Rev. Lett. \textbf{77}, 2360 (1996).

\bibitem{griffin}
A.~Griffin, Wen-Chin~Wu, and S.~Stringari, Phys. Rev. Lett. \textbf{78}, 1838
(1997).

\bibitem{minguzzi2}
M.~Amoruso \textit{et al.}, Eur. Phys. J. D \textbf{7}, 441 (1999). 

\bibitem{kagan}
Yu.~Kagan, E.~L.~Surkov, and G.~V.~Shlyapnikov, Phys. Rev. A \textbf{55}, R18
(1997).

\bibitem{baranov}
M.~A.~Baranov and D.~S.~Petrov, Phys. Rev. A \textbf{62}, 041601 (2000);
G.~M.~Bruun and C.~W.~Clark, Phys. Rev. Lett. \textbf{83}, 5415 (1999).

\bibitem{david scissors}
D.~Gu\'ery-Odelin and S.~Stringari, Phys. Rev. Lett. \textbf{83}, 4452 (1999).

\bibitem{foot}
O.~M.~Marag\`o \textit{et al.}, Phys. Rev. Lett. \textbf{84}, 2056 (2000). 

\bibitem{minguzzi}
A.~Minguzzi and M.~P.~Tosi, Phys. Rev. A \textbf{63}, 023609 (2001).

\bibitem{castin}
S.~Sinha and Y.~Castin, Phys. Rev. Lett. \textbf{87}, 190402 (2001).

\bibitem{chevy nucleation}
K.~W.~Madison \textit{et al.}, Phys. Rev. Lett. \textbf{86}, 4443 (2001).
One should not confuse the nucleation frequency with the critical frequency
required to ensure energetic stability to the vortex. This latter frequency is
significantly smaller.

\bibitem{recati}
A.~Recati, F.~Zambelli, and S.~Stringari, Phys. Rev. Lett. \textbf{86}, 377
(2001).

\bibitem{david}
D.~Gu\'ery-Odelin, Phys. Rev. A \textbf{62}, 033607 (2000).

\bibitem{cozzini}
M.~Cozzini and S.~Stringari, Phys. Rev. A \textbf{67}, 041602(R) (2003).
The hydrodynamic formalism developed in this paper refers to Bose-Einstein
condensates containing many vortical lines, where one can introduce the concept
of diffused vorticity.

\bibitem{zambelli}
F.~Zambelli, S.~Stringari, Phys. Rev. Lett. \textbf{81}, 1754 (1998).

\bibitem{chevy}
F.~Chevy, K.~W.~Madison, and J.~Dalibard, Phys. Rev. Lett. \textbf{85}, 2223
(2000).

\bibitem{viverit}
G.~M.~Bruun and L.~Viverit, Phys. Rev. A \textbf{64}, 063606 (2001);
G.~M.~Bruun and C.~W.~Clark, J. Phys. B \textbf{33}, 3953 (2000).

\bibitem{bulgac}
A.~Bulgac and Yongle Yu, cond-mat/0303235.

\bibitem{dalfovo}
F.~Dalfovo and S.~Stringari, Phys. Rev. A \textbf{63}, 011601 (2001).

\bibitem{ENS}
K.~W.~Madison \textit{et al.}, Phys. Rev. Lett. \textbf{84}, 806 (2000).

\end{thebibliography}
\end{document}